\documentclass[9pt,twocolumn,twoside]{osajnl}

\journal{pr} 

\setboolean{shortarticle}{false}

\title{Singlemode spatiotemporal soliton attractor in multimode GRIN fibers}

\author[1,*]{M. Zitelli}
\author[1]{M. Ferraro}
\author[2]{F. Mangini}
\author[1,3]{S. Wabnitz}

\affil[1]{Department of Information Engineering, Electronics and Telecommunications (DIET), Sapienza University of Rome, Via Eudossiana 18, 00184 Rome, Italy}
\affil[2]{Department of Information Engineering (DII), University of Brescia, Via Branze 38, 25123 Brescia, Italy}
\affil[3]{Novosibirsk State University, Pirogova 1, Novosibirsk 630090, Russia}

\affil[*]{Corresponding author: mario.zitelli@uniroma1.it}

\usepackage{hyperref}

\begin{abstract}
Experimental and numerical studies of spatiotemporal femtosecond soliton propagation over up to 1 km spans of parabolic graded-index (GRIN) fibers reveal that initial multimode soliton pulses naturally and irreversibly evolve into a singlemode soliton. This is carried by the fundamental mode of the fiber, which acts as a dynamical attractor of the multimode system for up to the record value (for multimode fibers) of 5600 chromatic dispersion distances.
This experimental evidence 
invalidates the use of variational approaches, which intrinsically require that the initial multimode propagation of a self-imaging soliton is indefinitely maintained.
 
\end{abstract}

\setboolean{displaycopyright}{true}

\begin{document}

\maketitle

\section{Introduction}

Optical solitons in fibers have been extensively and successfully studied over the past fifty years, leading to significant progress in long-distance optical communications and mode-locked lasers \cite{Hasegawabook,ZaWabook}. Although nearly all of these investigations involved the generation and propagation of singlemode fiber solitons, optical solitons can be supported by multimode optical fibers (MMFs) as well \cite{Hasegawa:80,Crosignani:81,Crosignani:82,doi:10.1063/1.5119434}. 

Interest in MMFs has been motivated by their potential for increasing the transmission capacity of long-distance optical links via the technique of mode-division-multiplexing (MDM), exploiting the multiple transverse modes of the fiber as information carriers \cite{Richardson}. In this context, it has been predicted that, in the presence of random mode coupling and nonlinearity, MMFs can support the stable propagation of Manakov solitons, leading to a nonlinear compensation of modal dispersion \cite{Mecozzi:12,2012arXiv1207.6506M}. The possibility of overcoming modal dispersion is also of great interest for high-speed local-area networks, where MMFs are extensively employed \cite{Agrawalbook}. In addition, there is a significant industrial interest in the use of large-area fibers for up-scaling the power of fiber lasers, for high-power beam delivery, and for biomedical imaging applications. In these applications, it is very important to maintain the high beam quality of singlemode fibers even when transporting a beam via an MMF \cite{Fermann:98,moussa2020spatiotemporal}.

When compared with singlemode fiber solitons, experimental studies of optical solitons in MMFs remain relatively scarce. The analysis of the propagation of femtosecond pulses undergoing soliton self-frequency shift (SSFS) in graded-inded (GRIN) MMFs has revealed that multiple transverse modes are effectively mutually trapped by cross-phase modulation, in spite of their strong linear temporal walk-off due to modal dispersion \cite{Renninger2012R31,Wright:15,Zhu:16,Zitelli:20}.   

To date, the dynamics and stability of MMF solitons remains yet to be assessed. Theoretical treatments of spatiotemporal soliton propagation in MMFs mostly rely on the variational approach (VA) \cite{Yu1995R35,RAGHAVAN2000377}. When neglecting the temporal dimension, the variational method permits to include the Kerr effect in the description of the periodic spatial self-imaging (SSI) which occurs in GRIN MMFs \cite{Karlsson:92}. Based on this VA, and by adding group-velocity dispersion, it has been proposed that soliton propagation in MMFs can be theoretically described in terms of a reduced one-dimensional generalized nonlinear Schr\"odinger equation (GNLSE) with a spatially varying effective mode area \cite{Conforti:17,Ahsan:18,Ahsan:19}. The validity of the VA is based on the assumption that the initial beam shape is maintained unchanged upon propagation, except for a limited set of slowly-varying parameters (e.g., the beam amplitude and width). 

Now, early experiments of soliton generation in GRIN MMFs have demonstrated that optical solitons generated by a highly multimode pump via the mechanism of Raman cascade and SSFS are essentially carried by the fundamental mode of the fiber \cite{Grudinin:88}. Although the mechanism of spatial beam reshaping for soliton pulses remained largely unexplained, it was attributed to a process of Raman beam-clean up. In recent years, new experiments revealed that the Kerr effect can also lead to spatial beam self-cleaning in GRIN MMFs. In this case, an irreversible transfer of energy towards the fundamental mode of the fiber is induced by modal four-wave mixing (FWM) processes, quasi-phase-matched by SSI \cite{Krupanatphotonics}. The occurrence of spatial beam self-cleaning has been investigated both in the normal \cite{Krupanatphotonics,LiuKerr,WrightNP2016} and anomalous dispersion regime of MMFs \cite{Leventoux:20}, but only in situations where temporal (chromatic and modal) dispersion does not play a role. 

In this work, we theoretically and experimentally demonstrate, we believe for the first time, the spatial beam self-cleaning of multimode solitons in GRIN MMFs. An initially excited multimode femtosecond soliton composed by a few low-order transverse modes, irreversibly decays into a singlemode soliton, owing to the FWM-induced energy transfer of higher-order modes into the fundamental mode of the MMF. Once formed, the singlemode soliton remained stable over the tested fiber length of 1 km, which corresponds to the record transmission distance in a MMF of 4600 modal dispersion distances, and 5600 chromatic dispersion lengths. This effect is of particular importance for technological applications, as it reveals that nonlinearity can counteract the effects of modal dispersion and random mode coupling, and enable the stable transport of high spatial quality beams over long distances by means of large area MMFs. As a side aspect, our results invalidate theoretical predictions based on the VA, since the beam shape substantially evolves along the fiber, in a way that the initial beam profile is not maintained.  

\section{Transmissions up to 1 km of GRIN fiber}
In this section, we provide a detailed description of the spatiotemporal evolution of multimode femtosecond solitons in long spans of parabolic GRIN fibers. Experimental characterizations of the output pulsewidth, bandwidth, and beam shape are supported by successful comparisons with extensive numerical simulations.
\subsection{Model and simulations}
A numerical model suitable for studying the propagation of multimodal pulses over long spans of GRIN fiber is based on the coupled-mode equations approach~\cite{Poletti:08}, which requires a preliminary knowledge of the input power distribution among the fiber modes. The model couples the propagating mode fields via Kerr and Raman nonlinearities, via four-wave mixing (FWM) terms of the type $Q_{plmn}A_lA_mA^*_n$; the coupling coefficients $Q_{plmn}$ are proportional to the overlap integrals of the transverse modal field distributions. 
We modified the standard coupled-mode generalized nonlinear Schr\"oedinger equations of Poletti and Horak~\cite{Poletti:08}, as implemented by Wright et al. as open source Matlab parallel numerical mode solver~\cite{8141863}, in order to include the wavelength-dependent linear losses of silica. Fiber dispersion and nonlinearity parameters are estimated to be $\beta_2=-28.8$ ps$^2$/km at 1550 nm, $\beta_3=0.142$ ps$^3$/km; nonlinear index $n_2=2.7\times 10^{-20}$ m$^2$/W, Raman response $h_r(\tau)$ with typical times of 12.2 and 32 fs~\cite{Stolen:89,Agrawal01}.

When considering a beam entering the fiber with no input tilt angle, and focused on the entry face with a 30 $\mu$m ($1\slash e^2$) diameter, by means a specific software we calculated the input mode relative power distribution to be: 52$\%$, 30$\%$, and 18$\%$ for the first 3 axial-symmetric modes $LP_{01}$, $LP_{02}$, $LP_{03}$ respectively; other higher-order modes carry negligible power. 
The sum over $l$, $m$, and $n$ of the FWM coupling coefficients $Q_{1lmn}$, $Q_{2lmn}$, $Q_{3lmn}$, responsible for feeding the 3 modes is $4.67\times 10^{9}$, $4.17\times 10^{9}$, $3.50\times 10^{9}$ m$^{-2}$, respectively; the number of coefficients larger than the mean value is 37, out of which 16 couple to the mode $LP_{01}$, 11 to the $LP_{02}$, 10 to the $LP_{03}$. The lack of symmetry between coupling coefficients is responsible for a slow, but irreversible transfer of energy from higher-order modes towards the fundamental, when pulses carried by different modes are temporally and spatially overlapping.


\begin{figure}[ht!]
\centering\includegraphics[width=8.8cm]{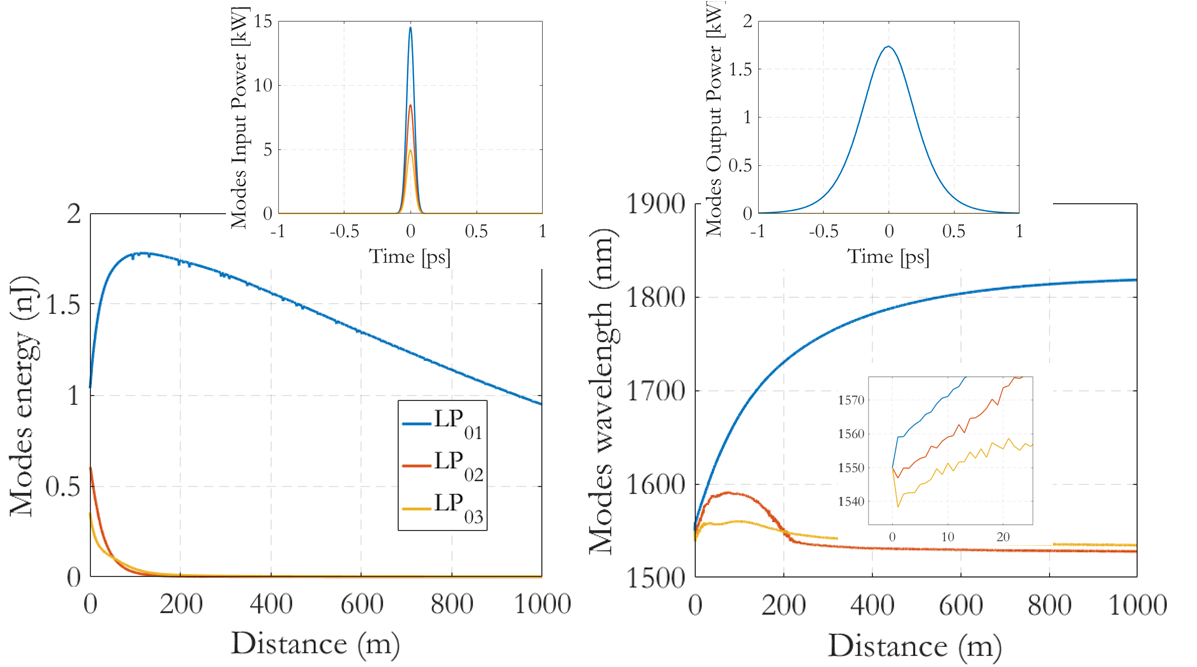}
\caption{Simulated energy and wavelength evolution of the three input axial modes. The two upper insets show the pulse power modal distribution at the input (left), and after 1 km of propagation (right).}
\label{Fig1}
\end{figure}

Fig.~\ref{Fig1} shows the simulated evolution along the GRIN fiber of the pulse energy in each of the three propagating modes (bottom-left), and their corresponding center wavelength (bottom-right). Here we coupled at the fiber input a Gaussian pulse with 67 fs duration, 1550 nm wavelength, 30 $\mu$m diameter (at $1\slash e^2$ of peak intensity), and 28 kW of peak power, which is suitable for spatiotemporal soliton generation. The top-left inset shows the input temporal power profile for each of the three modes. During propagation over the first 10 meters of fiber (see the wavelength panel in Fig.~\ref{Fig1}), the three modes separate their wavelengths, which permits them to experience different chromatic dispersions, which leads to a group-velocity difference that exactly compensates for modal dispersion walk-off. As a result, the three modes propagate together with the same speed: a spatiotemporal soliton is formed, which is characterized by the fact that its constituent modes remain mutually trapped in time. Remarkably, Fig.~\ref{Fig1} reveals that, as a result of nonlinear coupling between the fundamental mode and the other two higher-order axial modes, after approximately 120 m of propagation virtually all of the input energy is funneled into the pulse carried by the fundamental mode. Whereas higher-order modes decay into dispersive waves with negligible power, with a center wavelength close to the input value. The energy increase of the fundamental mode at the expense of higher-order modes can be approximated by an exponential law, specifically, $E_1 (z)=E_{in}[1-(1-f_1)e^{(-s_1 z)}]$, with $E_{in}$ the input energy, $f_1$=0.52 the initial power fraction of the fundamental mode, and $s_1$ a decay rate factor which depends on input pulse energy. 

Fig.~\ref{Fig1} shows that, in the subsequent 880 m of propagation, a substantially monomodal soliton propagates, experiencing progressive wavelength red-shift caused by Raman SSFS. As the soliton wavelength increases above 1700 nm, it starts losing energy because of linear fiber attenuation: as a result, it broadens temporally, in order to adiabatically conserve the monomodal solitonic condition (top-right inset).

It was previously shown analytically~\cite{Mecozzi:12,mumtaz2012nonlinear} that, in the case of degenerate modes with weak linear mode coupling, i.e., for dispersion and nonlinearity lengths which are much shorter than the random linear coupling length between modes, there is a solitonic solution for the scalar Manakov equations governing multimodal fiber transmission.

Here, numerical simulations reveal the presence of a more complex dynamics, where spatiotemporal solitons are formed from nondegenerate modes, over distances in the range of a few nonlinearity lengths $L_{NL}=\lambda w^2_e/(2n_2P_p)$ (here $w_e$ the effective beam waist and is $P_p$ the pulse peak power), provided that $L_{NL}$ is comparable with the modal walk-off length. In the example of Fig.~\ref{Fig1} and also in our experiments, after few meters distance a soliton with initial pulsewidth of 120 fs is generated, corresponding to a nonlinearity length of 18 cm, equal to the chromatic dispersion length; the modal walk-off length is approximately 22 cm, comparable to $L_{NL}$.

Once that the modes are temporally trapped, the spatiotemporal soliton experiences a Raman induced wavelength red-shift or SSFS.
As a result of SSFS, the pulsewidth of the fundamental soliton increases almost linearly with distance, because of the wavelength increase that leads to experiencing progressively larger (in absolute value) chromatic dispersions, and the consequent need to maintain the solitonic energy $E_1=\lambda|\beta_2(\lambda)|w_e^2\slash(n_2 T_0)$,
with $T_0=T_{\text{FWHM}}\slash1.763$.
The increase with distance of the soliton pulsewidth can be approximated by the law $T_0(z)=T_0(z_f)[1+s_2(z-z_f)]$, where $s_2$ is a slope which depends on input peak power, and $z_f$ is the initial distance of soliton formation. 


Simulations of Fig.~\ref{Fig1} predict that a slow power transfer from higher-order modes into the fundamental mode occurs upon propagation. This process is completed at distance in the range of hundreds or even thousands of nonlinearity lengths. After 100-150 m of propagation, the resulting soliton is substantially monomodal.

\subsection{Experimental evidence over 1 km of GRIN fiber}
In order to experimentally confirm the generation of monomodal solitons over long spans of GRIN multimodal fiber, a test-bed was prepared by using femtosecond pulses propagating over 1 km of fiber. The experimental setup used for the generation of MMF solitons consists of an ultrashort laser system, including a hybrid optical parametric amplifier (OPA) of white-light continuum (Lightconversion ORPHEUS-F), pumped by a femtosecond Yb-based laser (Lightconversion PHAROS-SP-HP), generating pulses at 100 kHz repetition rate with Gaussian beam shape ($M^2=1.3$); the pulse temporal shape at 1550 nm is nearly Gaussian, with 67 fs pulsewidth. The laser beam is focused by a 50 mm lens into the fiber with $1\slash e^2$ input diameter of approximately 30 µm. Laser pulses enter the fiber with peak powers ranging between 100 W and 500 kW, which is regulated by using an external attenuator. 
The used fiber is a 1 km span of commercial parabolic GRIN fiber, with core radius $r_c=25$ $\mu$m, cladding radius 62.5 $\mu$m, cladding index $n_{clad}$=1.444 at 1550 nm, relative index difference $\Delta=0.0103$. At the GRIN fiber output, a micro-lens focuses the near-field on an InGaAs camera (Hamamatsu C12741-03); a second lens focuses the beam into an optical spectrum analyzer (Yokogawa AQ6370D) with wavelength range 600-1700 nm, and into a real-time multiple octave spectrum analyzer (Fastlite Mozza), with range 1100-3000 nm. The output pulse temporal shape is inspected by using an infrared fast photodiode and an 8 GHz digital oscilloscope with 30 ps overall time response (Teledyne Lecroy WavePro 804HD), and an intensity autocorrelator (APE pulseCheck 50) with femtosecond resolution.

\begin{figure}[ht!]
\centering\includegraphics[width=8.5cm]{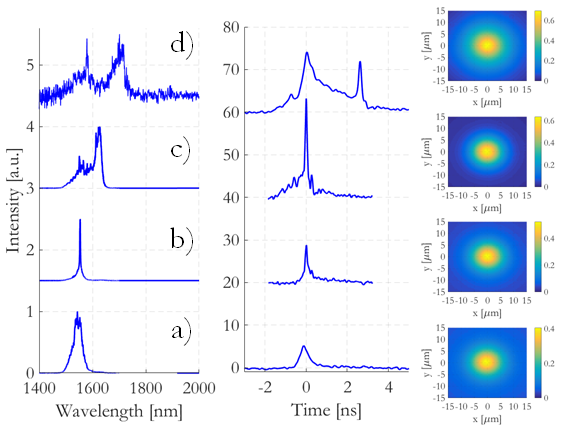}
\caption{Measured output spectra (left column), photodiode traces (center column), and near-field (right column) at 1 km distance, for input pulse peak powers of: a) 2 kW, b) 6.4 kW, c) 15 kW, d) 110 kW (see \href{Visualization 1.mp4}{Visualization 1}).}
\label{Fig3}
\end{figure}

Fig.~\ref{Fig3} shows an example of measured spectra (left column), photodiode traces (central column), and output near-fields (right column) after 1 km of fiber transmission, for input pulse peak powers of: a) 2 kW, b) 6.4 kW, c) 15 kW, d) 110 kW, respectively. At input peak power levels corresponding to the linear propagation regime (e.g., 2 kW), the output spectrum (see Fig.~\ref{Fig3}(a)) is nearly identical to that of the input pulse. Whereas the output pulse has temporally broadened to 1.2 ns, as a consequence of cumulated chromatic and modal dispersion. At 6.4 kW of peak power, an intermediate regime is observed, where self-phase modulation (SPM) nonlinearity starts counteracting dispersion, but it is not able to form a soliton yet: in the first few meters of propagation, the cumulated anomalous dispersion interacts with SPM, producing a rapid bandwidth compression and distortion of the chirped pulse (see Fig.~\ref{Fig3}(b))~\cite{Zitelli9306846}. At 15 kW of peak power, a spatiotemporal soliton starts to be formed (see Fig.~\ref{Fig3}(c)), and it experiences a Raman-induced SSFS~\cite{Mitschke:86}. As it can be seen, the recorded pulse is temporally narrower, and it is delayed with respect to the residual dispersive wave. In the soliton regime, the pulse bandwidth $\Delta\nu$ is not as narrow as in the intermediate regime; instead, it depends on the soliton pulsewidth as $\Delta\nu=0.315\slash T_{\text{FWHM}}$. For higher powers (110 kW, see Fig.~\ref{Fig3}(d)), a second and a third soliton is generated, each of them showing a similar behavior to the first one, and suffering Raman-induced time delay, according to the respective wavelengths.
The temporal evolution of the different solitons generated at the fiber output is better illustrated by the animation in \href{Visualization 1.mp4}{Visualization 1}, where the traces from the fast photodiode have been represented for increasing values of the input peak power.

With the help of numerical simulations, we predicted the formation of monomodal solitons after 100-150 m of propagation in a GRIN fiber. The experimental results demonstrate that more complex multimode soliton dynamics takes place when the input power grows substantially larger than the threshold for single soliton formation. Specifically, a train of monomodal solitons is generated across a wide range of input powers. 

\begin{figure}[ht!]
\centering\includegraphics[width=8.5cm]{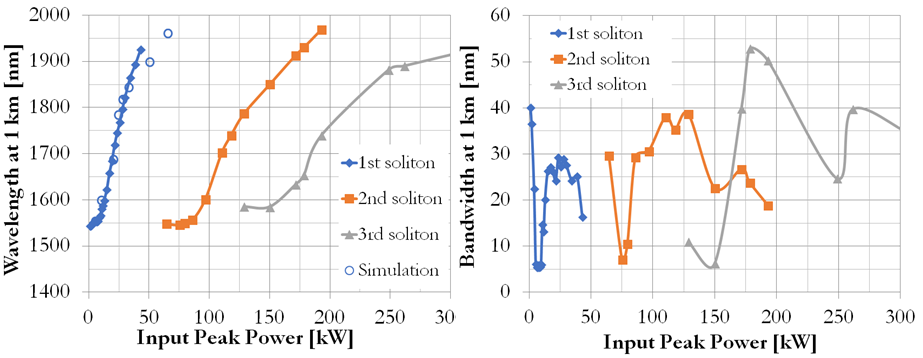}
\caption{Left: Measured wavelength for the three generated solitons vs. input peak power; Right: corresponding soliton bandwidth evolution.}
\label{Fig4}
\end{figure}

As it is illustrated in Fig.~\ref{Fig3}, recorded soliton spectra at 1 km distance are characterized by lobes with a sech shape: we measured their peak wavelength and bandwidth. The plots in Fig.~\ref{Fig4} (left) report the evolution of the output spectra vs. input peak power, and show that, for input powers between 10 kW and 43 kW, a first soliton is generated that experiences Raman SSFS. Whenever the wavelength of this soliton exceeds 1950 nm, it is absorbed by the fiber because of linear fiber loss. Numerical simulations (blue empty circles) report data which are in good agreement with the experiments. Small discrepancies for wavelengths longer than 1900 nm can be attributed to an over-estimation of the silica attenuation with respect to real fiber losses. For input peak powers between 60 kW and 190 kW, a second soliton is measured with a similar spectral shift, until a third soliton appears for input powers above 130 kW. The three solitons coexist in the fiber, but the first and second soliton are progressively absorbed when their wavelengths reach the 1900-2000 nm wavelength region. From Fig.~\ref{Fig4} (right), it can be confirmed that a strong reduction of the pulse bandwidth occurs, at input peak powers immediately below the value for soliton formation, as already discussed above.

As the soliton propagates through the parabolic GRIN fiber, its multimodal beam waist undergoes fast self-imaging (SI) oscillations~\cite{Hansson:20,PhysRevApplied.14.054063}. In this regime, the ratio of minimum to maximum beam effective area can be provided by the variational approach theory as~\cite{Karlsson:92,Ahsan:19} $C=\lambda^2 r_c^2\slash(2\pi^2 \Delta_{} n_{\text{eff}}^2 w_e^4)$ , with $r_c=25$ $\mu$m the core radius, $\Delta=0.0103$ the relative core-cladding index difference, and $n_{\text{eff}}=1.459$ the effective core index for the propagating mode. The effective beam waist $w_e$ is calculated for the fundamental mode by setting $C=1$, therefore $w_{e1}=(\lambda^2 r_c^2\slash2\pi^2 \Delta_{} n_{\text{eff}}^2)^{1\slash4}$, and equals $w_{e1}=7.7$ $\mu$m in our case.

\begin{figure}[ht!]
\centering\includegraphics[width=8.8cm]{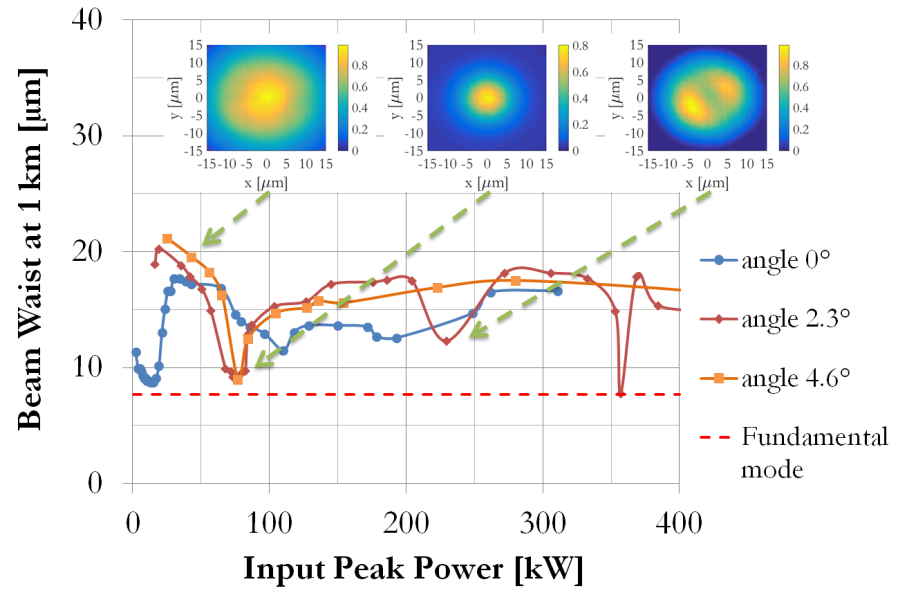}
\caption{Measured beam waist at 1 km distance vs. input peak power, when the input beam is coupled with 0$^\circ$, 2.3$^\circ$, and 4.6$^\circ$ tilt angle, as compared with the theoretical fundamental mode waist. Input beam waist is 15 $\mu$m. Insets show measured output beam shapes at powers indicated by the arrows.}
\label{Fig5}
\end{figure}

Fig.~\ref{Fig5} shows the measured output beam waist after 1 km of propagation, for increasing values of the input peak power; the beam size is compared with the theoretical monomodal value (horizontal dashed line). In order to enhance the beam cleaning effect, we coupled the input beam with different tilt angles, i.e., of 0$^\circ$, 2.3$^\circ$, and 4.6$^\circ$, respectively.

At a peak power of 15 kW and tilt 0$^\circ$, when the MMF solitonic pulse is formed and it starts experiencing Raman SSFS, the beam recorded at 1 km distance shows a strong diameter reduction down to the value of 8.5 $\mu$m, behaving as a substantially self-cleaned, monomodal soliton, as predicted by numerical simulations. For higher powers, the beam recovers a large waist: however, Fig.~\ref{Fig5} shows that the second and the third solitons also experience a beam width reduction (at 110 kW and at 190 kW, respectively), although their spatial compression is limited by the presence of multiple pulses and dispersive waves.

At an input tilt angle of 2.3°, we calculated that the power distribution between the modes at the input is $12\%$, $12\%$, $12\%$, $8.6\%$, $5.3\%$ for modes $LP_{01}$, $LP_{11e}$, $LP_{11o}$, $LP_{21}$, $LP_{02}$ respectively, and it also contains smaller proportions of higher-order modes. At 4.6$^\circ$ input angle, the power distribution is even more uniform. As a consequence, the generation of a fundamental soliton could still be measured at the fiber output, but for input powers that were increased by a factor of 5 with respect to the case of pure axial incidence. Still, a fundamental soliton was observed at 70-80 kW of input peak power, with the output beam waist of 8.8 $\mu$m, close to the value of the fundamental mode. It is interesting to note that the second and third solitons are still capable to produce virtually monomodal solitons with 2.3$^\circ$ input angle, demonstrating that the attraction property into the fundamental mode is not exclusive of the lowest-power soliton. 

The insets in Fig.~\ref{Fig5} show the recorded output near-fields when the input tilt angle is 2.3$^\circ$: the formation of a narrow, cleaned beam at the input power of 72 kW can be clearly seen, corresponding to the generation of a spatiotemporal fundamental mode soliton (central inset). This can be compared to the multimodal output beam which is obtained at low powers (16 kW, left inset). Whereas at higher input energies, the breakup of the soliton and the generation of dispersive waves produce wider beams, and in some cases higher-order mode patterns (230 kW, right inset). At the soliton power, we measured a beam quality factor $M^2=1.4$, which is close to the value $M^2=1.3$ of the input laser beam. 

\subsection{Experimental results over 120 m of GRIN fiber}
Numerical simulations in Section 2A have shown that, at a distance of approximately 120 m and at the solitonic power, nearly all of the pulse energy is nonlinearly transferred into the fundamental mode of the GRIN MMF (see Fig.~\ref{Fig1}). For longer distances, the wavelength shifted pulse starts suffering the effects of linear fiber loss. This observation led us to perform a soliton transmission experiment using a 120 m span of GRIN fiber, in order to observe and characterize the newly generated monomodal spatiotemporal soliton in the temporal, spatial, and frequency domains, respectively.

\begin{figure}[ht!]
\centering\includegraphics[width=8cm]{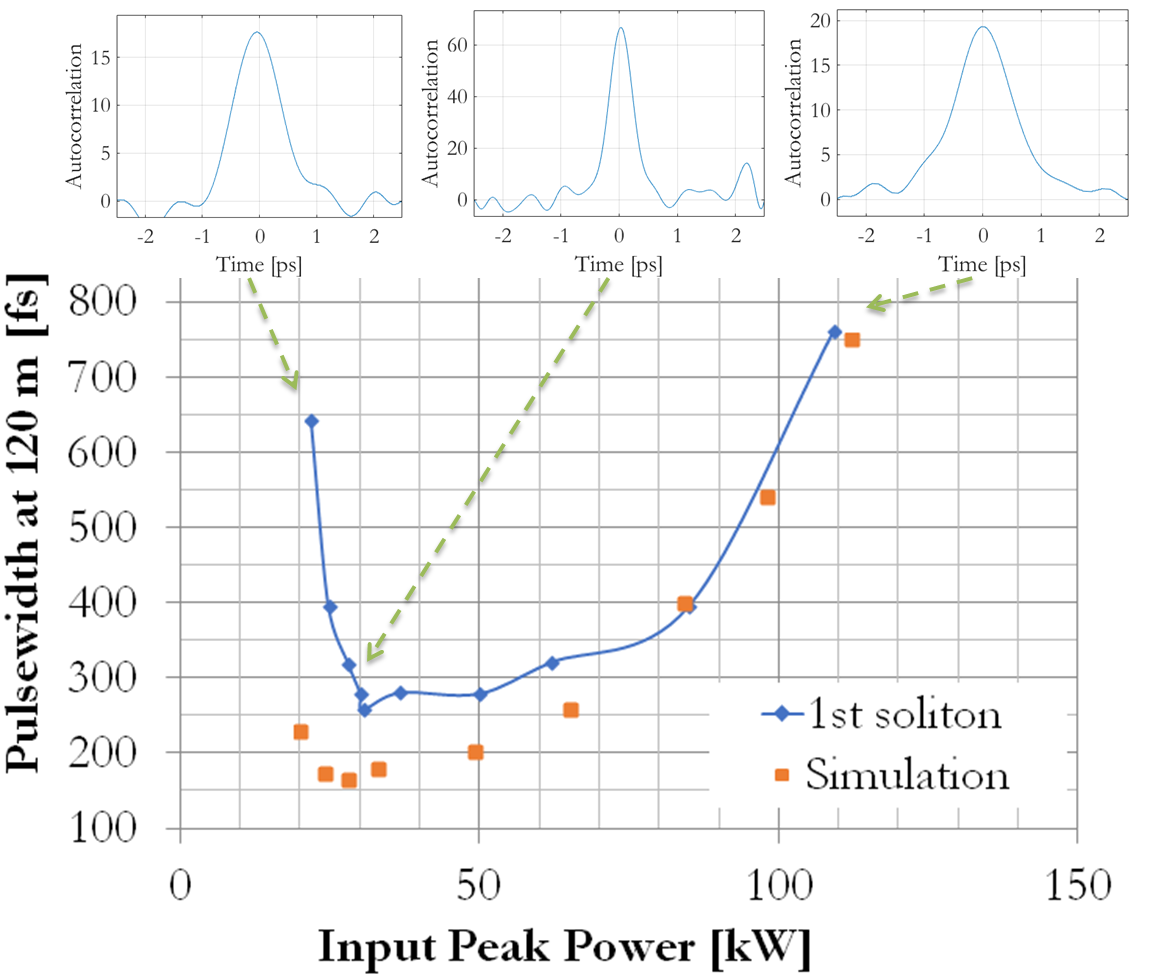}
\caption{Measured and simulated FWHM pulsewidth at 120 m distance vs. input peak power. The top insets are autocorrelation traces at 21 kW, 28 kW and 109 kW input power, respectively.}
\label{Fig6}
\end{figure}

Fig.~\ref{Fig6} shows the output pulsewidth of the soliton as it was measured by autocorrelations: experimental results are compared to simulations at 120 m. The top insets show autocorrelation traces at 21 kW, 28 kW, and 109 kW input power, respectively.
A minimum pulsewidth soliton was measured at the input peak power of 25-30 kW, with a temporal duration of $T_{\text{FWHM}}=260$ fs, against the simulated duration of 180 fs: the discrepancy may be attributed to dispersive effects from the output optics. Experiments and simulations performed with 10 m of fiber (not shown) provided output spatiotemporal solitons with 120 fs pulsewidth, and dispersion/nonlinearity length of 18 cm. The pulse width increases with distance or for higher input powers, because of the soliton wavelength increase induced by SSFS, so that it experiences a larger amount of anomalous chromatic dispersion. As a result, the pulse duration must grow larger in order to keep the fundamental soliton condition unchanged.


\begin{figure}[ht!]
\centering\includegraphics[width=8.5cm]{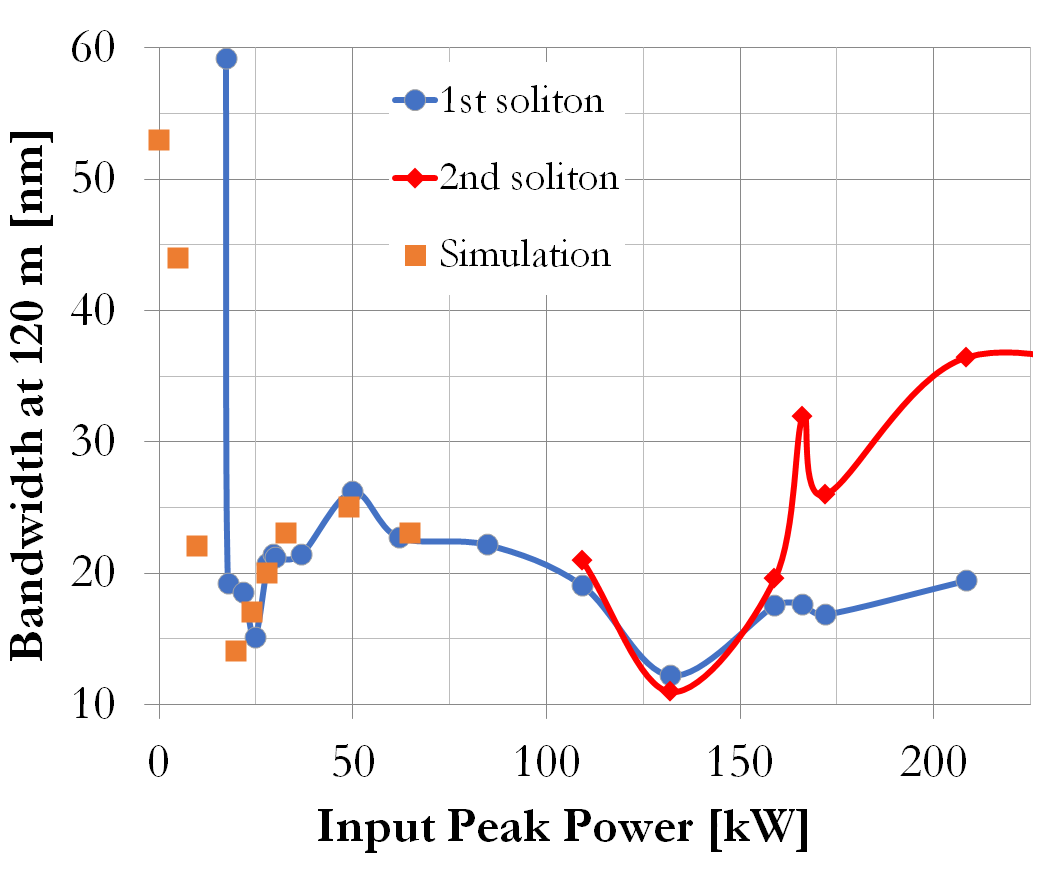}
\caption{Measured bandwidth vs. input peak power, for the two solitons observed after 120 m of GRIN fiber.}
\label{Sup1}
\end{figure}

Fig.~\ref{Sup1} shows the measured bandwidth of the spectral lobes corresponding to the first and second soliton, at 120 m distance; we also included the corresponding simulation results for the first soliton. At 15-20 kW of input peak power, the output pulse experiences a bandwidth compression down to 14 nm, before the phase of soliton formation. At 25-30 kW of input peak power, the first soliton has formed: the corresponding output bandwidth increases up to about 20 nm. 
Fig.~\ref{Sup1} also shows that the bandwidth of the second soliton, which is generated at input powers around 165 kW, exhibits a similar behavior. 


\begin{figure}[ht!]
\centering\includegraphics[width=7cm]{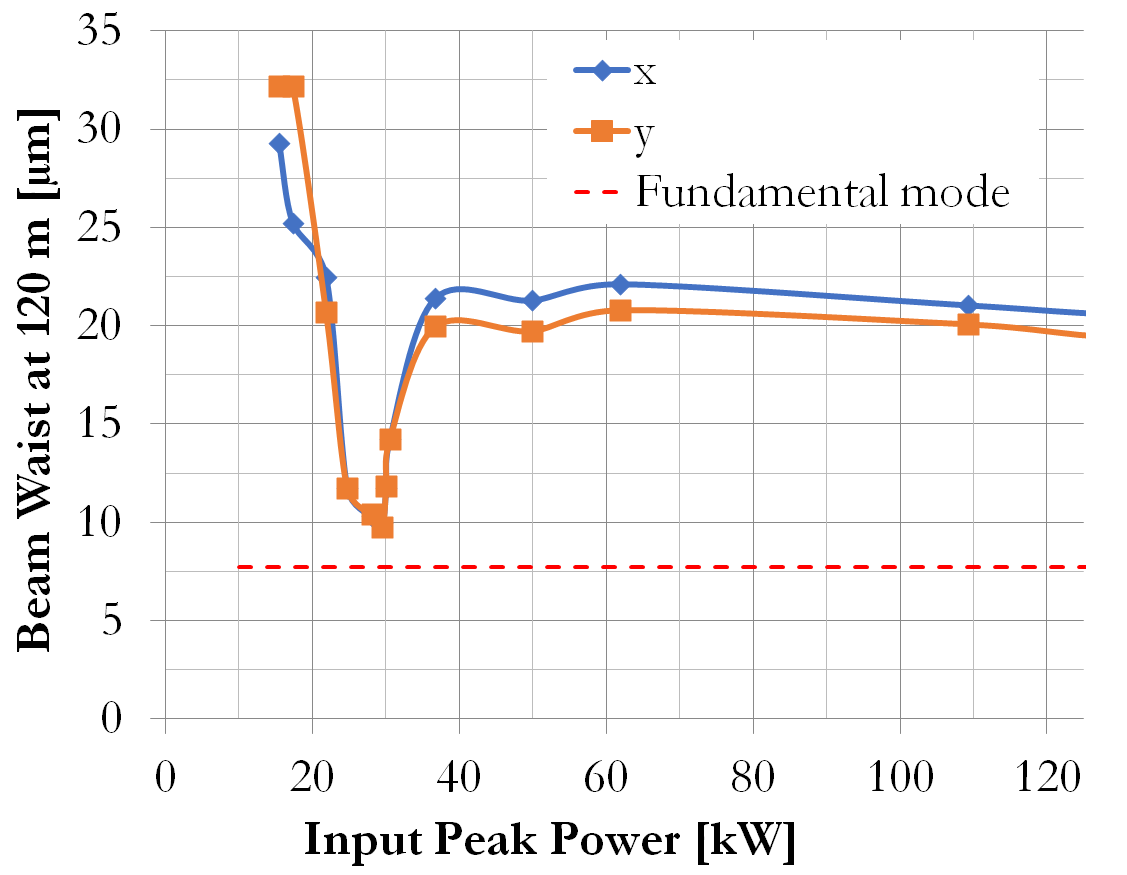}
\caption{Near field-beam waist vs. input peak power, for the first soliton at 120 m of GRIN fiber length.}
\label{Sup3}
\end{figure}
As far as the measured beam waist at 120 m is concerned, Fig.~\ref{Sup3} shows that, at 27-30 kW of input peak power, a minimum of output beam waist of 8.8 $\mu$m is reached, which again is close to the monomodal value of 7.7 $\mu$m. 

\begin{figure}[ht!]
\centering\includegraphics[width=8cm]{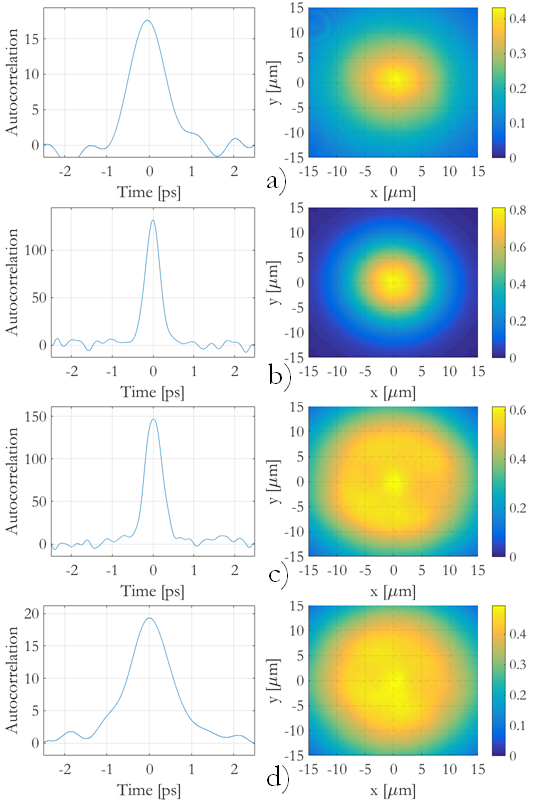}
\caption{Recorded autocorrelation traces (left column) and near-field beams (right column) from 120 m of GRIN fiber, for input peak powers: a) 22 kW, b) 29 kW, c) 37 kW, d) 109 kW.}
\label{Sup4}
\end{figure}

In Fig.~\ref{Sup4}, the recorded autocorrelation traces are given alongside with the corresponding near-field of the beam, for the input peak powers:~\ref{Sup4}.a) 22 kW,~\ref{Sup4}.b) 29 kW,~\ref{Sup4}.c) 37 kW, and ~\ref{Sup4}.d) 109 kW, respectively. As it can be seen, in the stage preceding soliton formation, at 22 kW of input peak power, the output beam is still relatively wide, while the pulse bandwidth reaches its minimum value (see Fig.~\ref{Sup1}). At the power of 29 kW, a spatiotemporal fundamental soliton is generated: pulse duration and beam width simultaneously reach their minima, i.e., 260 fs and 8.8 $\mu$m, respectively, whereas the pulse bandwidth starts to increase. At 37 kW of input peak power, the output pulsewidth is still limited to 280 fs; to the contrary, the beam is no longer confined to a single mode, and the waist has more than doubled with respect to the fundamental soliton case; still, we may talk of a multimodal, spatiotemporal soliton in this case. For higher powers (109 kW) the pulse starts broadening in time, and the multimodal soliton condition is gradually lost.

Finally, from the simulated and measured data we can calculate the soliton order at both 120 m and 1 km of GRIN fiber length, by using the formula $N=[n_2 T_0 E_1/(\lambda |\beta_2(\lambda)|w_e^2)]^{1/2}$
By using the fiber dispersion value at the soliton wavelength, the simulated output energy and pulse width, and the effective waist $w_{e1}=7.7$ $\mu$m obtained from the variational theory, we obtain that, for a solitonic input peak power of 28 kW, the order is $N=1.17$ at 120 m, and $N=1.03$ at 1 km. Those results indicate that, in spite of the linear losses, a fundamental soliton of order 1 was reached at 1 km distance. Whereas at 120 m the effective waist to be used in the formula
to obtain $N=1$ is $w_e=8.8$ $\mu$m, which is exactly the value of the measured beam waist  (see Fig.~\ref{Sup3}).

\section{Conclusion}
Spatiotemporal soliton evolution in parabolic GRIN fibers was previously described, by using the variational approach, as a stable propagation of a self-imaging beam; in this picture,
the several modes formed at the input trap each other and indefinitely preserve their temporal shape.
In this work we demonstrated, both experimentally and numerically, that the spatiotemporal soliton evolution over long spans of fiber is far more complex: a pump pulse feeds a spatiotemporal soliton at proper input energy when the nonlinearity length of the forming soliton is comparable to the modal walk-off length; the modes effectively trap each other in time but, in the range of hundreds of nonlinearity lengths, a slow and irreversible energy transfer is observed from higher-order modes into the fundamental mode of the fiber, which acts as a dynamical attractor of the multimode system. The spatiotemporal soliton thus naturally evolves into a singlemode soliton, and it permanently maintains this state.
The optimal energy where the singlemode, minimum waist beam is observed at the fiber output corresponds to that of minimum output pulsewidth. The pulse bandwidth suffers a bandwidth reduction previous to the optimal solitonic power, and increases to the solitonic value at optimal energy. The output soliton order results unitary if the effective beam waist is considered equal to the singlemode value.

To place our findings in a broader perspective, we anticipate that our results will be of general fundamental interest, because they provide the first example of fully spatiotemporal beam condensation in classical nonlinear wave systems~\cite{PRA11b}. From the point of view of technological applications, the generation of a robust ultrashort soliton attractor in a nonlinear multimode fiber is of significance for the delivery of high energy laser beams in a variety of industrial applications, for high-power spatiotemporal mode-locked multimode fiber lasers~\cite{WiseScience}, and for the use of multimode fibers in high-bit-rate fiber optic networks.  

\section*{Acknowledgments}
We thank Wright et al. for making freely available the open source parallel numerical mode solver for the coupled-mode nonlinear Schr\"oedinger equations~\cite{8141863}, and Oleg Sidelnikov for developing a mode decomposition numerical code. 
We acknowledge the financial support from the European Research Council Advanced Grants No. 874596 and No. 740355 (STEMS), the Italian Ministry of University and Research (R18SPB8227), and the Russian Ministry of Science and Education Grant No. 14.Y26.31.0017.


\section*{Disclosures}
The authors declare no conflicts of interest.

\medskip

\bibliography{biblio}

\begin{thebibliography}{10}
\newcommand{\enquote}[1]{``#1''}

\bibitem{Hasegawabook}
A.~Hasegawa and M.~Matsumoto, \emph{Optical Solitons in Fibers}
  (Springer-Verlag Berlin Heidelberg, 2003).

\bibitem{ZaWabook}
V.~E. Zakharov and S.~Wabnitz, \emph{Optical Solitons: Theoretical Challenges
  and Industrial Perspectives} (Springer-Verlag Berlin Heidelberg, 1999).

\bibitem{Hasegawa:80}
A.~Hasegawa, \enquote{Self-confinement of multimode optical pulse in a glass
  fiber,} {\protect\JournalTitle{Opt. Lett.}} \textbf{5}, 416--417 (1980).

\bibitem{Crosignani:81}
B.~Crosignani and P.~D. Porto, \enquote{Soliton propagation in multimode
  optical fibers,} {\protect\JournalTitle{Opt. Lett.}} \textbf{6}, 329--330
  (1981).

\bibitem{Crosignani:82}
B.~Crosignani, A.~Cutolo, and P.~D. Porto, \enquote{Coupled-mode theory of
  nonlinear propagation in multimode and single-mode fibers: envelope solitons
  and self-confinement,} {\protect\JournalTitle{J. Opt. Soc. Am.}} \textbf{72},
  1136--1141 (1982).

\bibitem{doi:10.1063/1.5119434}
K.~Krupa, A.~Tonello, A.~Barth\'{e}l\'{e}my, T.~Mansuryan, V.~Couderc,
  G.~Millot, P.~Grelu, D.~Modotto, S.~A. Babin, and S.~Wabnitz,
  \enquote{Multimode nonlinear fiber optics, a spatiotemporal avenue,}
  {\protect\JournalTitle{APL Photonics}} \textbf{4}, 110901 (2019).

\bibitem{Richardson}
D.~J. Richardson, J.~M. Fini, and L.~Nelson, \enquote{Space-division
  multiplexing in optical fibres,} {\protect\JournalTitle{Nat. Photonics}}
  \textbf{7}, 354--362 (2013).

\bibitem{Mecozzi:12}
A.~Mecozzi, C.~Antonelli, and M.~Shtaif, \enquote{Coupled {M}anakov equations
  in multimode fibers with strongly coupled groups of modes,}
  {\protect\JournalTitle{Opt. Express}} \textbf{20}, 23436--23441 (2012).

\bibitem{2012arXiv1207.6506M}
A.~{Mecozzi}, C.~{Antonelli}, and M.~{Shtaif}, \enquote{{Soliton trapping in
  multimode fibers with random mode coupling},} {\protect\JournalTitle{arXiv
  e-prints}} arXiv:1207.6506 (2012).

\bibitem{Agrawalbook}
G.~Agrawal, \emph{Fiber-Optic Communication Systems} (Wiley-Interscience,
  1997).

\bibitem{Fermann:98}
M.~E. Fermann, \enquote{Single-mode excitation of multimode fibers with
  ultrashort pulses,} {\protect\JournalTitle{Opt. Lett.}} \textbf{23}, 52--54
  (1998).

\bibitem{moussa2020spatiotemporal}
N.~O. Moussa, T.~Mansuryan, C.~H. Hage, M.~Fabert, K.~Krupa, A.~Tonello,
  M.~Ferraro, L.~Leggio, M.~Zitelli, F.~Mangini, A.~Niang, G.~Millot, M.~Papi,
  S.~Wabnitz, and V.~Couderc, \enquote{Spatiotemporal beam self-cleaning for
  high-resolution nonlinear fluorescence imaging with multimode fibres,}
  (2020).

\bibitem{Renninger2012R31}
W.~H. Renninger and F.~W. Wise, \enquote{Optical solitons in graded-index
  multimode fibres,} {\protect\JournalTitle{Nat. Commun.}} \textbf{4}, 1719
  (2013).

\bibitem{Wright:15}
L.~G. Wright, W.~H. Renninger, D.~N. Christodoulides, and F.~W. Wise,
  \enquote{Spatiotemporal dynamics of multimode optical solitons,}
  {\protect\JournalTitle{Opt. Express}} \textbf{23}, 3492--3506 (2015).

\bibitem{Zhu:16}
Z.~Zhu, L.~G. Wright, D.~N. Christodoulides, and F.~W. Wise,
  \enquote{Observation of multimode solitons in few-mode fiber,}
  {\protect\JournalTitle{Opt. Lett.}} \textbf{41}, 4819--4822 (2016).

\bibitem{Zitelli:20}
M.~Zitelli, F.~Mangini, M.~Ferraro, A.~Niang, D.~Kharenko, and S.~Wabnitz,
  \enquote{High-energy soliton fission dynamics in multimode grin fiber,}
  {\protect\JournalTitle{Opt. Express}} \textbf{28}, 20473--20488 (2020).

\bibitem{Yu1995R35}
S.-S. Yu, C.-H. Chien, Y.~Lai, and J.~Wang, \enquote{Spatio-temporal solitary
  pulses in graded-index materials with {K}err nonlinearity,}
  {\protect\JournalTitle{Opt. Commun.}} \textbf{119}, 167--170 (1995).

\bibitem{RAGHAVAN2000377}
S.~Raghavan and G.~P. Agrawal, \enquote{Spatiotemporal solitons in
  inhomogeneous nonlinear media,} {\protect\JournalTitle{Optics
  Communications}} \textbf{180}, 377 -- 382 (2000).

\bibitem{Karlsson:92}
M.~Karlsson, D.~Anderson, and M.~Desaix, \enquote{Dynamics of self-focusing and
  self-phase modulation in a parabolic index optical fiber,}
  {\protect\JournalTitle{Opt. Lett.}} \textbf{17}, 22--24 (1992).

\bibitem{Conforti:17}
M.~Conforti, C.~M. Arabi, A.~Mussot, and A.~Kudlinski, \enquote{Fast and
  accurate modeling of nonlinear pulse propagation in graded-index multimode
  fibers,} {\protect\JournalTitle{Opt. Lett.}} \textbf{42}, 4004--4007 (2017).

\bibitem{Ahsan:18}
A.~S. Ahsan and G.~P. Agrawal, \enquote{Graded-index solitons in multimode
  fibers,} {\protect\JournalTitle{Opt. Lett.}} \textbf{43}, 3345--3348 (2018).

\bibitem{Ahsan:19}
A.~S. Ahsan and G.~P. Agrawal, \enquote{Spatio-temporal enhancement of
  raman-induced frequency shifts in graded-index multimode fibers,}
  {\protect\JournalTitle{Opt. Lett.}} \textbf{44}, 2637--2640 (2019).

\bibitem{Grudinin:88}
A.~B. Grudinin, E.~Dianov, D.~Korbkin, M.~P. A, and D.~Khaidarov,
  \enquote{Nonlinear mode coupling in multimode optical fibers; excitation of
  femtosecond-range stimulated-{R}aman-scattering solitons,}
  {\protect\JournalTitle{J. Exp. Theor. Phys. Lett}} \textbf{47}, 356--359
  (1988).

\bibitem{Krupanatphotonics}
K.~Krupa, A.~Tonello, B.~M. Shalaby, M.~Fabert, A.~Barth{\'e}l{\'e}my,
  G.~Millot, S.~Wabnitz, and V.~Couderc, \enquote{Spatial beam self-cleaning in
  multimode fibres,} {\protect\JournalTitle{Nat. Photonics}} \textbf{11},
  234--241 (2017).

\bibitem{LiuKerr}
Z.~Liu, L.~G. Wright, D.~N. Christodoulides, and F.~W. Wise, \enquote{Kerr
  self-cleaning of femtosecond-pulsed beams in graded-index multimode fiber,}
  {\protect\JournalTitle{Opt. Lett.}} \textbf{41}, 3675--3678 (2016).

\bibitem{WrightNP2016}
L.~G. Wright, Z.~Liu, D.~A. Nolan, M.-J. Li, D.~N. Christodoulides, and F.~W.
  Wise, \enquote{Self-organized instability in graded-index multimode fibres,}
  {\protect\JournalTitle{Nat. Photonics}} \textbf{10}, 771--776 (2016).

\bibitem{Leventoux:20}
Y.~Leventoux, A.~Parriaux, O.~Sidelnikov, G.~Granger, M.~Jossent, L.~Lavoute,
  D.~Gaponov, M.~Fabert, A.~Tonello, K.~Krupa, A.~Desfarges-Berthelemot,
  V.~Kermene, G.~Millot, S.~F\'{e}vrier, S.~Wabnitz, and V.~Couderc,
  \enquote{Highly efficient few-mode spatial beam self-cleaning at 1.5 $\mu$m,}
  {\protect\JournalTitle{Opt. Express}} \textbf{28}, 14333--14344 (2020).

\bibitem{Poletti:08}
F.~Poletti and P.~Horak, \enquote{Description of ultrashort pulse propagation
  in multimode optical fibers,} {\protect\JournalTitle{J. Opt. Soc. Am. B}}
  \textbf{25}, 1645--1654 (2008).

\bibitem{8141863}
L.~G. {Wright}, Z.~M. {Ziegler}, P.~M. {Lushnikov}, Z.~{Zhu}, M.~A. {Eftekhar},
  D.~N. {Christodoulides}, and F.~W. {Wise}, \enquote{Multimode nonlinear fiber
  optics: Massively parallel numerical solver, tutorial, and outlook,}
  {\protect\JournalTitle{IEEE Journal of Selected Topics in Quantum
  Electronics}} \textbf{24}, 1--16 (2018).

\bibitem{Stolen:89}
R.~H. Stolen, J.~P. Gordon, W.~J. Tomlinson, and H.~A. Haus, \enquote{Raman
  response function of silica-core fibers,} {\protect\JournalTitle{J. Opt. Soc.
  Am. B}} \textbf{6}, 1159--1166 (1989).

\bibitem{Agrawal01}
G.~P. Agrawal, \emph{Nonlinear Fiber Optics} (Third edition, Par. 2.3, Academic
  Pre, 2001).

\bibitem{mumtaz2012nonlinear}
S.~Mumtaz, R.-J. Essiambre, and G.~P. Agrawal, \enquote{Nonlinear propagation
  in multimode and multicore fibers: generalization of the manakov equations,}
  {\protect\JournalTitle{Journal of Lightwave Technology}} \textbf{31},
  398--406 (2012).

\bibitem{Zitelli9306846}
M.~{Zitelli}, M.~{Ferraro}, F.~{Mangini}, and S.~{Wabnitz}, \enquote{Managing
  self-phase modulation in pseudo-linear multimodal and monomodal systems,}
  {\protect\JournalTitle{Journal of Lightwave Technology}} pp. 1--1 (2020).

\bibitem{Mitschke:86}
F.~M. Mitschke and L.~F. Mollenauer, \enquote{Discovery of the soliton
  self-frequency shift,} {\protect\JournalTitle{Opt. Lett.}} \textbf{11},
  659--661 (1986).

\bibitem{Hansson:20}
T.~Hansson, A.~Tonello, T.~Mansuryan, F.~Mangini, M.~Zitelli, M.~Ferraro,
  A.~Niang, R.~Crescenzi, S.~Wabnitz, and V.~Couderc, \enquote{Nonlinear beam
  self-imaging and self-focusing dynamics in a grin multimode optical fiber:
  theory and experiments,} {\protect\JournalTitle{Opt. Express}} \textbf{28},
  24005--24021 (2020).

\bibitem{PhysRevApplied.14.054063}
F.~Mangini, M.~Ferraro, M.~Zitelli, A.~Niang, A.~Tonello, V.~Couderc, and
  S.~Wabnitz, \enquote{Multiphoton-absorption-excited up-conversion
  luminescence in optical fibers,} {\protect\JournalTitle{Phys. Rev. Applied}}
  \textbf{14}, 054063 (2020).

\bibitem{PRA11b}
P.~Aschieri, J.~Garnier, C.~Michel, V.~Doya, and A.~Picozzi,
  \enquote{Condensation and thermalization of classsical optical waves in a
  waveguide,} {\protect\JournalTitle{Phys. Rev. A}} \textbf{83}, 033838 (2011).

\bibitem{WiseScience}
L.~G. Wright, D.~N. Christodoulides, and F.~W. Wise, \enquote{Spatiotemporal
  mode-locking in multimode fiber lasers,} {\protect\JournalTitle{Science}}
  \textbf{358}, 94--97 (2017).

\end{thebibliography}

\end{document}